\documentstyle[emulateapj]{article}  

\def\asca{{\it ASCA}~}  
\def\cd{{\cal{D}}}  
    
\def\Del{\Delta}

\def\gam{\gamma}  
\def\gmax{\gamma_{\rm max}}  
\def\gmin{\gamma_{\rm min}}

\def\gok{\gamma_{_{\rm K}}}  
\def\kal{K${\alpha}~$}

\def\lk{l_{_{\rm K}}}

\def\mbh{M_{\bullet}}

\def\ok{\Omega_{_{\rm K}}}

\def\rhoss{\rho_{_{\rm SS}}}  
\def\sunm{M_{\odot}}  
\def\st{\sigma_{_{\rm Th}}}  
  
\def\tny{T_{_{\rm ADAF}}}  
\def\tss{T_{\rm ss}}

\slugcomment{To Appear in The Astrophysical Journal Letters}  
  
\begin{document}  
  
\title{Blob ejection from advection-dominated accretion flow:\\   
observational consequences}   
  
\author{Jian-Min Wang\altaffilmark{1},   
        Ye-Fei Yuan\altaffilmark{2},  
	Mei Wu\altaffilmark{1}, and  
        Masaaki Kusunose\altaffilmark{3}}  
  
\altaffiltext{1}{Laboratory of Cosmic Ray and High Energy Astrophysics, 
Institute of High Energy Physics, CAS,  
Beijing 100039, and CAS-PKU Beijing Astrophysical Center, Beijing  
100871, P.R. China, E-mail: wangjm@astrosv1.ihep.ac.cn}  
  
\altaffiltext{2}{Center for Astrophysics, USTC, Hefei 230026,   
and National Astronomical Observatories, CAS, P.R. China,   
E-mail: yfyuan@ustc.edu.cn}  
  
\altaffiltext{3}{Department of Physics, School of Science,  
Kwansei Gakuin University, Nishinomiya 662-8501, Japan,   
E-mail: kusunose@kwansei.ac.jp}  
  
\begin{abstract}  
There is increasing evidence for the presence of an optically thin   
advection-dominated accretion flow (ADAF) in low luminosity active galactic  
nuclei and radio-loud quasars. The present paper is devoted to explore  
the fate of a blob ejected from an ADAF, and to discuss its observational  
consequences. It is inevitable for the ejected blob to drastically expand  
into its surroundings. Consequently, it is expected that a group of  
relativistic electrons should be accelerated, which may lead to nonthermal  
flares, since a strong shock will be formed by the interaction between the  
blob and its surroundings. Then the blob cools down efficiently, leading to  
the appearance of recombination lines about $10^5$s after its ejection from  
an ADAF.  We apply this model to NGC 4258 for some observational prediction, 
and to PKS 2149--306 for the explanation of observational evidence. 
Future simultaneous observations of recombination X-ray lines and continuum  
emission are highly desired to test the present model.  
   
\subjectheadings{accretion: accretion disk - jet: emission - galaxies: nuclei}  
\end{abstract}  
  
\section{Introduction} 
The process of accretion onto a black hole is thought to be an efficient way  
to release gravitational energy. However for low enough accretion  
rates the released gravitational energy around a black hole is stored as  
entropy of accreting gas rather than being radiated. It is then expected that  
the internal energy of accreting gas will be very high and the temperature may  
reach the virial value. This is the well-known optically thin ADAF (Narayan  
\& Yi 1994), which is quite different from that of standard accretion disks  
(Shakura \& Sunyaev 1973).  
  
Recently, based on the standard model of accretion disks, Tagger \& Pellat  
(1999) found an accretion-ejection instability. The essential physics for 
this instability is that the extracted angular momentum is not emitted  
radially but ends up as an Alfv\'{e}n wave which can eject material in the  
direction perpendicular to the disk plane.  Such an  
instability may be stronger in ADAFs, because the instability grows with 
disk thickness although the detail of the instability in 
ADAFs remains open (Tagger 2000, private communication).  
Such a process provides another potential channel to release the  
gravitational energy in black hole accreting systems.  The interesting 
properties of \kal line emission in radio-loud quasars lead to 
the suggestion that the central engine of radio-loud quasars may be 
powered by optically thin ADAFs (Sambruna et al. 2000a, b). 
If such an instability works in an ADAF,  
some blobs should be ejected. Then a natural question arises:  
what is the fate of a blob ejected from the optically thin ADAF? Comparison  
of the energy density of the ejected blob with that of the inter-cloud medium  
(ICM) in broad line regions (BLR) will give a clue to this question. The 
typical values of density and temperature of the ICM in the BLR are thought 
to be $5.0\times 10^7$K and $5.0\times 10^6$ cm$^{-3}$, respectively,  
in radio-loud quasars (Netzer 1991).  If a blob is  
ejected from an ADAF, what are its observational consequences both in line  
and continuum emission?  These are the main goals of the present paper.  
  
\section{The Fate of a Blob Ejected from an ADAF}  
The physical properties of the ejected blob should be obtained from the global 
solution of an ADAF, which has already been done by Narayan et al. (1997).  
Since we are primarily interested in the fate of a blob, we can avoid the 
details of global structure of an ADAF.  For simplicity, we deal  
with the two limiting cases to constrain the flow, i.e., fully and partially  
advected structure with free-free emission. The actual disk temperature and 
density functions in practice will be somewhere in between those given by  
these two limiting cases. 
  
\subsection{Optically Thin ADAF}  
For the first limiting case, namely that of full advection, we use the 
self-similar solution of Narayan \& Yi (1994), and get temperature  
$ \tny \approx 2\times 10^{12}r^{-1}$K, and density  $n_{_{\rm ADAF}}   
\approx 1.1 \times 10^{11}\alpha^{-1}\dot{m}M_8^{-1}r^{-1.5}$cm$^{-3}$,   
where $\alpha$ is the viscosity parameter, $\dot{m}$  
is the dimensionless accretion rate by 
$\dot{M}_{\rm Edd}=1.4\times 10^{26}M_8$(g/s)(the accretion efficient 
$\eta=0.1$), $M_8$ is the mass of the central black  
hole in units of $10^8\sunm$, and $r=R/R_g$ is the dimensionless 
radius with $R_g=2G\mbh/c^2$ and $\mbh$ is the mass of black hole. 
  
We consider the other limiting case of an accretion disk,   
assuming a standard accretion disk (Shakura \& Sunyaev 1973) holds. 
In such a case the flow cools via free-free emission.  
In this work we only consider the consequences of 
a single-temperature treatment as a first approximation. 
The height-averaged equations of an accretion disk  
are used since a more sophisticated approach is not warranted for this 
level of treatment.  Using the  
$\alpha$-viscosity law, the dissipated energy production rate  
$Q^+=\alpha p \ok\gok$, where $\gok=-d \ln \ok/d \ln R$  
and $\ok=(G\mbh/R^3)^{1/2}$ is the Keplerian angular velocity,  
is balanced by the free-free cooling rate 
$Q^-=\Lambda_0\rho^2T^{1/2}$ ($\Lambda_0=1.24\times 10^{21}$). 
Suppose that the vertical hydrostatic  
balance holds, i.e., $H^2=c_s^2/\ok^2=kT/m_p\ok^2$, 
where $H$ is the half thickness of a disk, $k$ is the Boltzmann constant,  
and $m_p$ is the proton mass.  The $\phi$-direction equation reads  
$\dot{M}\lk=4\pi \alpha R^2pH$, where $\lk$ is the Keplerian angular momentum. 
For simplicity, we have neglected the inner boundary condition 
that no heat generation occurs at the inner boundary. 
This is a reasonable approximation for the region 
not very close to the inner radius.   
Temperature and density are then given by  
\begin{equation}  
\tss=8.3\times 10^{11}r^{-3/4}\alpha^{-1}\dot{m}^{1/2}~~{\rm K} \, ,  
\end{equation}  
\begin{equation}  
n_{\rm ss}=3.47\times 10^{13} r^{-15/8} \alpha^{1/2}\dot{m}^{1/4} M_8^{-1}~~  
       {\rm cm^{-3}} \, .  
\end{equation}  
Then the actual temperature $T$ and number density $n$ should be  
$\tss \le T \le \tny$ and $n_{_{\rm ADAF}} \le n \le n_{\rm ss}$. 
We can see that even the free-free cooling disk is of much higher   
temperature (within $10^4 R_g$) than the one  
producing recombination lines.   
  
\subsection{Blob Ejected from Optically Thin ADAF}  
In ADAFs the magnetic field strength is close to the equipartition value   
(Narayan \& Yi 1995).  The self-similar solution for such magnetized ADAFs 
has been found by Soria et al. (1997).  It is interesting to note that there  
is an accretion-ejection instability within several gravitational radii in   
magnetized geometrically thin disks (Tagger \& Pellat 1999).   
Such an accretion-ejection instability in ADAFs would be stronger than that  
in thin disks because its growth rate will increase with increasing disk  
thickness (Tagger 2000, private communication), though the instability in  
ADAFs needs to be studied in detail in future.  
Since most of the dissipated energy will be mainly deposited on protons,  
the internal energy density ($U=\case{3}{2} n_{\rm ss}k\tss$)  
of accreting gas can be obtained by equations (1) and (2): 
\begin{equation}  
U= 5.9\times 10^9 r^{-\frac{21}{8}}  
   \alpha^{-\frac{1}{2}}\dot{m}^{\frac{3}{4}}M_8^{-1}  
                 ~~{\rm erg/cm^3} \, .  
\end{equation}  
We can see from above equation that the internal energy density is very 
sensitive to the ejection location of a disk, i.e., $U\propto r^{-21/8}$.   
Thus the ejection at large radii is less significant. It is then assumed  
that a blob with size $R_0(\le R_e)$ is ejected from an ADAF at radius  
$R_e$ via such an instability, and that the most likely ejection radius is  
within $R_e=10R_g$. Also we assume that changes in temperature and density 
are little during the ejection. 
  
The ejected blob will expand rapidly, and strongly interact with its   
surroundings. This behavior is very similar to the fireball of $\gam$-ray  
bursts (Piran 1999).  The interaction between the blob and its surroundings   
is very complex.  There are two reasons for this complexity, because of two  
kinds of interactions between the ejected blob and its surrounding medium,  
i.e., expansion and transverse motions in the medium.  In this paper we  
pay our main attention to the interaction between the blob expansion and it   
surrounding medium.  Neglecting the detailed process of expansion (Sedov 1969)  
due to the high internal energy of the blob, we have 
$d(\rhoss v^2/2)/dt=-dU/dt$, for the expansion velocity, which is based on 
the energy conservation. Similarly to the fireball of $\gam$-ray bursts, the 
expansion will be decelerated when the swept mass by expansion is equal to 
the blob initial mass  $M_{\rm blob}$, namely, 
$\case{4}{3}\pi R_2^3 n_2m_p=M_{\rm blob}$,  
where $R_2$ is its final radius, $R_2=R_0+v\tau \approx v\tau$,  
with $\tau$ being the timescale of expansion.  These equations give  
$v=(3kT_1/m_p)^{1/2}$ and $\tau=(R_0/v) (n_1/n_2)^{1/3}$,  
where $T_1$ and $n_1$ are the temperature and the number density of the ejected  
blob at the initial state, respectively. The expansion  proceeds with Mach  
number  
\begin{equation}  
\label{eq:mach} 
{\cal M}^2\approx \left(\frac{v}{c_{s2}}\right)^2 \approx \frac{T_1}{T_2},  
\end{equation}  
where $c_{s2}$ is the sound speed of its surroundings.  Typically   
${\cal M}^2 \sim 2\times 10^3$, for $T_2=5.0\times 10^{7}$K and $T_1=10^{11}$K,  
and strong shock will be formed.  It is expected that the temperature of the  
post shock wave will go down to $T_1/{\cal M}^2$ (Sedov 1969). Thus the ejected  
blob will be efficiently cooled down, so that the condition for some  
recombination lines, e.g., FeXXV or FeXXIV ions, will be satisfied (Raymond  
\& Smith 1977). The separation of the ejected blob from the central black hole  
would be $d=\beta c\tau$, where $\beta c$ is the ejection velocity of the blob.  
There are mainly three effects strongly affecting the observed   
profile of emission lines: 1) spherical expansion causes the broadening of  
emission line profile, 2) the gravitational redshift (perhaps not strong),   
and 3) Doppler shift due to the bulk motion of the blob.  The profile of an 
emission line from the relativistic outflow close to a rotating black hole has 
been calculated in our recent paper (Wang et al. 2000).  It is expected that   
a line from such an ejected blob will have a more complex profile.  
The evolution of the blob will be mainly  
determined by free-free cooling  after its swept mass of the surrounding  
medium equals to the initial mass of the blob.  The density is then  
approximated as   
$n_e\approx \left[n_1R_0^3+n_2(v\tau)^3\right]/(v\tau)^3=2n_2=1.0\times 10^7$  
(cm$^{-3}$).  The timescale is $\tau_{\rm ff}\approx 6.6T_8^{1/2}n^{-1}_{e,7}$ 
(yr), where $T_8=T/10^8$ and $n_{e,7}=n_e/10^7$.  It is thus expected that the  
blob with temperature $\sim 10^8$K will cool down at 2pc from the nucleus, if  
the ejection occurs at almost light speed.  
  
Next we briefly describe the possibility of nonthermal emission from a blob; 
blast waves are expected to be generated during expansion and  
to accelerate a group of thermal electrons to be relativistic.  
The conversion of the blob kinetic energy to nonthermal particles  
involves complicated plasma physics and shock acceleration processes.   
However, we can avoid dealing with the detailed acceleration physics by simply 
assuming that some fraction of the internal energy is transformed into a group 
of nonthermal electrons (e.g., Blandford \& Eichler 1987). Here we assume that 
fraction $\xi$ of all the internal energy is channelled into the nonthermal 
electrons. From the energy conservation, the total number of relativistic 
electrons can be estimated from 
$\xi R_0^3(n_1k\tss)=(v\tau)^3n_0m_ec^2\int \gam^{1-p}d\gam$,  
where we assumed that shock acceleration forms a population of electrons with  
power-law index $p$ as $n_0\gam^{-p}$. The total number of relativistic  
electrons in one blob is  
\begin{equation}  
N_e=\frac{4}{3}\pi \xi n_1R_0^3\left(\frac{k\tss}{m_ec^2}\right)  
               \left(\frac{2-p}{1-p}\right)  
               \left(\frac{\gmax^{1-p}-\gmin^{1-p}}  
               {\gmax^{2-p}-\gmin^{2-p}}\right).  
\end{equation}  
The fraction of the relativistic electrons to the total number of the initial  
blob is about $\xi k\tss/\gmin m_ec^2=0.012\xi_{-2}$ for $p>2$, where   
$\xi_{-2}=\xi/0.01$ and $\gmin=50$. This strongly implies that the radiating  
electrons may originate from the present mechanism, which may be a proposing  
mechanism  responsible for the acceleration of relativistic electrons in some  
of radio-loud quasars. It is expected the nonthermal emission spectrum 
and light curves 
should be similar to some properties predicted by Li \& Kusunose (2000) and 
Kusunose et al. (2000). 
  
It is thus inevitable for the ejected blob that the high-temperature drops  
due to the strong interaction between the blob and its surroundings, and as a  
natural consequence there is a group of electrons to be accelerated to  
relativistic energy responsible for the continuum and some recombination lines  
from the cooled blob.  
 
\section{Applications}  
It has been generally accepted that accretion onto super massive black holes  
leads to the release of gravitational energy in active galactic nuclei 
(Rees 1984). However, the status of the accretion disk in different kinds of 
AGNs remains uncertain.  Ion-supported tori may power the central engine in  
radio-loud quasars (Rees et al. 1982).  A famous low luminosity AGN NGC 4258  
is generally thought as representative of ADAFs (Gammie et al. 1999). However, 
recent observations of \kal  by \asca show the origination is not from an 
accretion disk (Reynolds et al. 2000).  Yaqoob et al. (1999) detected a highly  
Doppler blue-shifted K-emission line in PKS 2149--306.  We thus apply our  
model to the two objects.  
  
{\it NGC 4258:} The mass of the central black hole is well determined to be  
$\mbh=3.6\times 10^7\sunm$ by Miyoshi et al. (1995).  Radio observation set  
the transition radius from the standard disk to the ADAF to be 100$R_g$  
(Herrnstein et al. 1999), and the accretion rate is  
$\sim 1.0\times 10^{-3}\dot{M}_{\rm Edd}$ (Gammie et al. 1999).  
Taking the typical value of $\alpha=0.1$,  
we have $T_1=4.7\times 10^{10}$K and $n_1=7.2\times 10^{10}$cm$^{-3}$,  
if the ejection takes place at 10$R_g$, where the viscous heating rate is 
highest. The initial dimension of  
the ejected blob is taken to be $R_0=10R_g=10^{14}$ cm.  The additional  
component with luminosity $L_X^A=2.0\times 10^{40}$ ergs/s in X-ray shows  
that the tenuous plasma is of temperature 0.5 keV (Reynolds et al. 2000). 
This sets constraints on the density of the environment in the vicinity of 
the nucleus via the Thomson scattering depth $\tau_{_{\rm Th}} = \st n_2 \ell$,
where $\st$ is the Thomson scattering cross section  and $\ell$ is the 
dimension of its surroundings. Using the X-ray luminosity of the additional 
component contributed from the tenuous plasma  
$L_X^A=\frac{4}{3}\pi\ell^3 \Lambda n_2^2T_e^{1/2}$ 
($\Lambda=3.4\times 10^{-27}$), 
we have the number density of the surroundings to be  
$n_2=10^8(\tau_{_{\rm Th}}/0.1)^3$cm$^{-3}$. From  
equation (4) we have the Mach number ${\cal M} = 97$. 
The timescale of cooling due to expansion is then given by  
$\tau=2.36\times 10^5$s, which is much shorter than that of free-free cooling.   
The expanded radius of the blob is $R_2=9.0\times 10^{14}$cm. The temperature  
will drop from $4.7\times 10^{10}$K to  
$T=T_1/{\cal M}^2\approx 5.0\times 10^6$K. During the cooling some  
recombination lines will be produced (Raymond \& Smith 1977), such as   
FeXXVI$\lambda1.78$\AA, FeXXV$\lambda1.85$\AA, FeXXV$\lambda1.86$\AA, and  
oxygen line OVIII$\lambda18.97$, and silicon Si XIV$\lambda 6.18$\AA, etc.  
If the blob moves with Doppler factor of 10 (Gammie et al. 1999),  
we expect to observe the highly Doppler   
blue-shifted line at 64$\cd_{10}$ keV, where $\cd=10\cd_{10}$ is the  
Doppler factor.  Such lines should be detected by future observations of  
{\it INTEGRAL}.   
  
{\it PKS 2149--306:} The highly Doppler shifted iron K-emission line   
has been detected by \asca in this high redshift ($z$=2.345)  
radio-loud quasar (Yaqoob et al. 1999), although the current data does not  
allow to determine its unambiguous profile.  We attempt to apply our present  
model to this source. The central mass can be obtained from the 
full-width-at-half-maximum (FWHM) and the luminosity of emission line CIV   
(Peterson 1998).  With the measurements of $v_{_{\rm FWHM}}=6400$km/s  
(Wilkes 1986), and $L({\rm CIV})=1.2\times 10^{46}$erg/s (Wilkes et al. 1983)  
(which we measured from the spectrum), we have the mass of the central black hole,  
$\mbh=3.4\times 10^9\sunm$, which corresponds to Eddington luminosity  
$L_{_{\rm Edd}}=4.3\times 10^{47}$erg/s.  If we follow the mean spectrum of   
radio-loud quasars (Elvis et al. 1994), we get the bolometric luminosity    
$L_{\rm bol}\approx 5.0\times 10^{47}$erg/s based on the continuum   
(Siebert et al. 1996).  Then the accretion rate is roughly  
$ \dot{m} = L_{\rm bol}/(\cd^4L_{_{\rm Edd}}) \approx 2.0\times 10^{-2}$,  
where Doppler factor $\cd=2.65$; here we directly used the Doppler factor of  
the iron K emission line as jet's. It suggests that this object may be powered  
by an ADAF. The transition radius from the standard disk to the ADAF can be 
obtained by the approximate formula  
$r_{\rm tr}\approx 2.6\times 10^3 \alpha^4\dot{m}^{-2}$ for the case that   
half the released energy is advected [their eq. (4.1) of Narayan \& Yi 1995].  
We get $r_{\rm tr}\approx 400$ for $\alpha=0.1$.  If the ejection takes place  
at $10R_g$ with initial radius $R_0=10R_g\approx 10^{16}$cm, then the density   
is $n_1=1.7\times 10^8$cm$^{-3}$ and the temperature is 
$T_1=2.3\times 10^{11}$K.   
Taking the typical values of the parameters, $n_2=5.0\times 10^6$cm$^{-3}$ and 
$T_2=5.0\times 10^7$K, for ICM in the BLR (Netzer 1991), we have the typical  
expansion timescale $\tau \approx 2.8\times 10^6$s, and the temperature will  
drop to $10^{7\sim 8}$K, where Mach number ${\cal M}^2=4.7\times 10^3$ from  
equation (4).  The blob's distance from the center is about  
$\beta c\tau \approx 8.4\times 10^{16}$cm$\approx 80R_g$ ($\beta\approx 1$).  
It is thus expected that some high energy recombination lines will appear in   
$2.8\times 10^6$s since the ejection of a blob.  
 
There should be two steps to eject a blob. The first is the blob  
formation with the timescale of $\tau_{\rm f}$, which is approximate to  
the viscous timescale (timescale to accumulate matter), namely  
$\tau_{\rm f} \approx \alpha^{-1}\tau_{_{\rm K}}$  
[$\tau_{_{\rm K}}=1/\ok=(R_g/c)r^{3/2}=10^6$s at $r=10$]. In the more accurate 
global solution the radial velocity is higher than the above one. The second  
is the relativistic ejection with timescale $\tau_{\rm e}$.  
It seems reasonable to assume  
that the timescale of ejection is much shorter than the timescale of blob  
formation.  The duration for a blob to produce 
the iron recombination lines is approximately 
$\tau_{\rm ff}=6.6$yr for $T_e=10^8$K and $n_e=1.0\times 10^7$cm$^{-3}$.  
Therefore the number of the blobs producing the recombination line of iron  
K-emission would be $N_b\approx \tau_{\rm ff}/(\tau_{\rm e}+\tau_{\rm f}) 
\approx \alpha \tau_{\rm ff}/\tau_{_{\rm K}} \approx 20$ if the ejection takes  
place at 10$R_g$ (It should be noted that this number is {\it not} the total  
of the ejected blobs). We assume the cosmic abundance of iron in the blob.  
Considering the boosting effects of flux due to the relativistic motion,   
the intrinsic flux is given by $F_{\rm in}=F_{\rm obs}\cd_b^{-4}$,   
where $F_{\rm obs}$ is the observed flux (erg~s$^{-1}$~cm$^{-2}$)  
and $\cd_b$ is the Doppler factor of outflow. The energy of K-emission line  
with emissivity $j_{_{\rm K}}$ within the time interval $\Del t$ is   
$\Del E\approx \frac{4}{3}\pi N_b j_{_{\rm K}} n_e^2 R_2^3 \Del t$,   
which can be compared with the observable  
$\Del E=4\pi d_L^2F_{\rm obs}\Del t/\cd_b^4$,   
where $n_e$ is the number density of electrons in the blob.  With the  
help of energy conservation, we can predict the observed flux:  
$F_{\rm obs}=N_{\rm b}\cd_b^4 j_{_{\rm K}}n_e^2 R_2^3/3d_L^2 
\approx 0.8\times 10^{-13}$erg~s$^{-1}$~cm$^{-2}$, where we take the  
emissivity of iron recombination lines,  
$j_{_{\rm K}}$=1.4$\times 10^{-24}$erg~s$^{-1}$~cm$^{-3}$ 
(Raymond \& Smith 1977), and $\cd_b=2.65$ (Yaqoob et al 1999). We find this  
estimated value is close to the observed flux, 
$F_{\rm obs} \approx 3.6^{+2.5}_{-2.5}\times 10^{-13}$erg~s$^{-1}$~cm$^{-2}$ 
(Yaqoob et al. 1999). Here we do not rule out the contribution of iron 
\kal fluorescence emission to the observed flux. The total number of 
relativistic electrons in one blob is $N_e\approx 1.5\times 10^{54}$, 
if we take the minimum energy of electrons $\gmin=50$.  These electrons will 
be responsible for the nonthermal emission.  The detailed radiation processes  
and spectra with the evolution of electron's energy 
distribution will be carried out in a separate paper.  
  
\section{Conclusions and Discussions}  
If a blob ejection takes place in such an ADAF the fate of the blob  
is an interesting topic. In the ADAF regime the released gravitational energy 
from the viscous dissipation is restored as the entropy of the accreting gas.   
We have explored the observational consequences of the blob ejected from an 
optically thin ADAF. We found it inevitably drastically expands into its 
surroundings with Mach number of ${\cal M} \approx 10^2$. The hot blob cools 
down rapidly via the interaction between the blob and its surroundings via 
the blob's expansion. Consequently a group of thermal electrons will be 
accelerated to be relativistic via strong shocks, leading to nonthermal 
radiation from the blob. Simultaneously there will appear some of recombination
lines from the blob. If the ejected blob moves, these lines will be highly 
Doppler shifted. We think the observed iron K emission line in PKS 2149--306 
results from  the ejected blob from an optically thin ADAF. We also made 
predictions for NGC 4258 for future observations of {\it INTEGRAL}.   
  
Recent studies on the \kal line properties in radio-loud quasars show that  
the line is weak and narrow in radio-loud quasars, but strong and broad in   
radio-quiet quasars, which are quite similar to that in Seyfert galaxies  
(Reynolds 1997, Sambruna et al. 1999a, Wozniak et al. 1998, Reeves \&  
Turner 2000, Fabian et al. 2000). The ADAF thus may power the central  
engines of radio-loud quasars (Sambruna et al. 1999b). The proposed model is  
consistent with the existing data of PKS 2149-306 and 
supported by the recent observations of \kal line in other radio-loud    
quasars. The thermal X-ray emission lines from an ADAF have been studied by  
Narayan \& Raymond (1999), who suggest these lines may be detected by  
the current X-ray instruments (Narayan \& Raymond 1999, Perna et al. 2000).  
If the future observations (high space resolution of radio telescopes and  
high energy resolution of X-ray telescopes) confirm our predictions, the  
highly Doppler blue-shifted iron K emission line may shed light on jet  
formation and be very helpful to the construction of central engine models  
of radio-loud quasars.  
 
It is likely for the ejected blob to encounter BLR clouds when it passes 
through the BLR. One may expect to see changes in the optical/UV emission 
line spectrum in the simultaneous multiwavelength observations. For a 
radio-loud quasar it would be difficult to test the X-ray continuum from 
the ADAF itself since the nonthermal emission dominates its continuum  
by the Doppler boosting effects. The spatial distribution of ejected
blobs is expected to be perpendicular to the accretion disk plane.
Their distributions may lead to interesting observables, for example,
the line profile will be broadened,  if they occupy a considerable
solid angle. 
 
The proposed model is based on two important assumptions. First, we assumed 
a blob is ejected from an ADAF, but we did not specify the working mechanism 
responsible for the ejection. This leads us not to accurately predict the 
number of blobs. The accretion-ejection instability found by Tagger \& Pellat 
(1999) could be a promising mechanism and we will study it in future.   
The second involves the interaction between the blob with its surroundings.  
If the density of blob's surroundings is too low,  
the fate of the ejected blob would be different from the proposed model.    
 
\acknowledgements{The authors are grateful to an anonymous referee for  
the helpful comments and suggestions on the manuscript clarifying several 
points. The useful discussions with T.P. Li, F.J. Lu and J.L. Qu are 
acknowledged. J.M.W. is supported by 'Hundred Talents Program of CAS'. 
This work is financed by the Special Funds for Major State Basic Research 
Projects and by NSFC.}

\end{document}